# Effect of Zn substitution on morphology and magnetic properties of copper ferrite nanofibers

Weiwei Pan, Fengmei Gu, Kuo Qi, Qingfang Liu, Jianbo Wang [a]

*Institute of Applied Magnetics, Key Laboratory of Magnetism and Magnetic Materials of Ministry of Education, Lanzhou University, Lanzhou 730000, People's Republic of China*

**Abstract:**

Spinel ferrite $Cu_{1-x}Zn_xFe_2O_4$ nanofibers over a compositional range $0 < x < 1$ were prepared by electrospinning combined with sol-gel method. The influence of $Zn^{2+}$ ions substitution on morphology, structure, and magnetic properties of copper ferrite has been investigated. The results show that surface of $CuFe_2O_4$ nanofibers consists of small open porosity, while surface of doped nanofibers reveals smooth and densified nature. With increasing Zn substitution, saturation magnetization initially increases and then decreases with a maximum value of 58.4 emu/g at $x = 0.4$, coercivity and square ratio all decrease. The influence of substitution on magnetic properties is related with the cation distraction and exchange interactions between spinel lattices.

a)  Electronic mail: **wangjb@lzu.edu.cn**

Tel: +86-0931-8914171 and Fax: +86-0931-8914160

# 1. Introduction

Interesting in spinel ferrites ($M$Fe$_2$O$_4$, $M$ = Co, Ni, Mn, Mg, Zn, etc) has greatly increased in the past few years due to their remarkable magnetic, electrical, optical properties and compelling applications in many areas [1-3]. In particular, nanostructure spinel ferrites have gained considerable attention owing to their physical properties are quite different from those of bulk and particle [4-5]. Among various nanostructures, nanofibers are a promising candidate as building blocks in nanoscale sensors, nanophotonics, nanocomputers, and so forth. Spinel ferrites nanofibers therefore have become an important subject of many research groups. Electrospinning provides a versatile, low cost, and simple process for fabricating nanofibers with diameters in the range from several micrometers down to tens of nanometers. It has been studied and patented in 1940s by Formhals [6], and it has been used to synthesize a variety of materials, including organic composites, inorganic composites, and ceramic magnetic materials from now. NiFe$_2$O$_4$, CoFe$_2$O$_4$, MgFe$_2$O$_4$, and MnFe$_2$O$_4$ ferrites nanofibers have been synthesized by electrospinning and their structural and magnetic properties are investigated in detail [7-10].

Copper ferrite (CuFe$_2$O$_4$) is an interesting material and has been widely used for various applications, such as catalysts for environment [11], gas sensor [12], and hydrogen production [13]. Magnetic and electrical properties of spinel ferrites vary greatly with the change chemical component and cation distribution. For instance, most of bulk CuFe$_2$O$_4$ has an inverse spinel structure, with 85% Cu$^{2+}$ occuping B sites [14], whereas ZnFe$_2$O$_4$ is usually assumed to be a completely normal spinel with all $Fe^{3+}$ ions on B sites and all $Zn^{2+}$ ions on A sites [15]. $Zn^{2+}$ ions occupy preferentially A sites while $Fe^{3+}$ ions would be displaced from A sites for B sites. Zn-substitution results to a change of cations in chemical composition and a different distribution of cations between A and B sites. Consequently the magnetic and electrical properties of spinel ferrites will change with changing cation distribution. The influence of Zn-substitution on magnetic and structure of $Co_{1-x}Zn_xFe_2O_4$ and $Ni_{1-x}Zn_xFe_2O_4$ nanofibers have been reported [16-18]. However, there are no reports on the synthesis and characterize of Cu-Zn ferrites nanofibers in the literatures.

In this paper, we successfully prepared $Cu_{1-x}Zn_xFe_2O_4$ ferrites nanofibers via electrospinning technique combined with sol-gel, and the influence of $Zn^{2+}$ substitution on structure and magnetic properties of ferrite nanofibers was investigated. Magnetic state of spinel ferrite nanofibers changes from ferrimagnetic to paramagnetic was observed.

## 2. Experimental

In a typical procedure, a appropriate amount of copper nitrate ($Cu(NO_3)_2 \cdot 3H_2O$), zinc nitrate ($Zn(NO_3)_2 \cdot 6H_2O$), and ferric nitrate ($Fe(NO_3)_3 \cdot 9H_2O$) with 1-x:x:2 molar ratios of Cu:Zn:Fe were dissolved in *N,N*-Dimetylformamide (DMF), followed by magnetic stirring for 3 h to ensure the completely dissolution of metal salt. Simultaneously, poly (vinyl pyrrolidone) (PVP, $M_W \approx 1, 300, 000,$) was dissolved in ethanol (1 mL) and magnetic stirred for 3 h. Then 1 mL metal nitrates/DMF solution was added slowly to the PVP/ethanol solution under continuous stirred for 2 h with PVP concentration of 8 wt% for electrospinning. The obtained electrospinning solution was loaded in a plastic syringe with a stainless steel needle, which was connected to high-voltage equipment. A piece of aluminum foil used as the collector was placed in front of the needle tip as the negative electrode. The applied voltage was 12.5 kV and the distance between syringe needle tip and collector was 14 cm. The collected as-spun $Cu_{1-x}Zn_xFe_2O_4$ nanofibers were dried at 50 $^o$C for 5h. The dried nanofibers subsequently were calcined at 650 $^o$C for 3 h under an ambient atmosphere with heating rate of 1 $^o$C/min.

X-ray diffraction (XRD) patterns were recorded by a PANalytical diffractometer using Cu$K\alpha$ radiation with $\lambda$ = 0.15418 nm. Morphology, diameter, and chemical composition of calcined nanofibers were characterized by field emission scanning electron microscopy (FE-SEM, Hitachi S-4800) and Energy dispersive spectroscopy (EDS). Magnetization properties measurement of the calcined nanofibers were carried out on a vibrating sample magnetometer (VSM, Lakeshore 7403, USA) with a maximum applied field of 12 kOe at room temperature.

**3. Results and discussion**

3.1 XRD analysis

Figure 1 shows XRD patterns of $Cu_{1-x}Zn_xFe_2O_4$ (x = 0.0, 0.2, 0.4, 0.6, 0.8, and 1.0) ferrites nanofibers , which were prepared by electrospinning at room temperature. Lattice parameter (*a*), average grain size (*D*) and X-ray density ($\rho_x$) are listed in Table 1. XRD patterns show that all peaks indexed to pure cubic phase, where (220), (311), (400), (422), (511), and (440) represent the main crystal phase in $CoFe_2O_4$ spinel ferrite. As shown in inset of Figure 1, the position of peaks shifts slightly to lower scattering angles in the series $Cu_{1-x}Zn_xFe_2O_4$ nanofibers in accordance with the slight increase of the lattice parameter (Figure 2). This can be predominantly attributed to the replacement of smaller $Cu^{2+}$ ions (0.72 Å) by larger $Zn^{2+}$ ions (0.74 Å). The X-ray density ($\rho_x$) was calculated according to following relation; $\rho_x = ZM/Na^3$, where, Z is the number of molecules per unit cell (Z = 8), $M$ is the molecular weight, N represent the Avogadro's number and $a$ is the lattice parameter of the ferrite [19]. Figure 2 shows that as zinc content increases, X-ray density increases linearly. It directly correlates with the increase of molecular weight. The average grain size $D$ were calculated from x-ray line broadening of the (311), (511), (440) diffraction peaks using Scherrer equation. The value of $D$ varies from 26.1 nm ($CuFe_2O_4$) to 32.1 nm ($Cu_{0.2}Zn_{0.8}Fe_2O_4$), which denotes that zinc substitution leads to an increase of average grain size of $Cu_{1-x}Zn_xFe_2O_4$ nanofibers.

3.2 Morphological and componential analysis

Figure 3 shows representative SEM images of $Cu_{1-x}Zn_xFe_2O_4$ nanofibers with different Zn content calcined at 650 °C for 3 h in air. The surface of $CuFe_2O_4$ nanofibers is relatively rough, with a small quantity of open porosity (as shown in Figure 3 (a)). When Zn content is increases, smooth surface and densified microstructure of nanofibers are predominant. The diameter of $Cu_{1-x}Zn_xFe_2O_4$ nanofibers dependent on zinc content is presented in Figure 4. It can be seen that diameters of $CuFe_2O_4$, $Cu_{0.8}Zn_{0.2}Fe_2O_4$ and $ZnFe_2O_4$ nanofibers are about 110, 80, and 60 nm, respectively, clearly indicating $Zn^{2+}$ substitution leads to an obvious decrease of diameter. EDS is used to confirm chemical composition of $Cu_{1-x}Zn_xFe_2O_4$ nanofibers as shown in Figure 5, detail data is given in Table 2. The number of Fe element do not change much with increasing zinc content, but the numbers of Cu and Zn element change obviously with zinc content. For instance, $Cu_{0.6}Zn_{0.4}Fe_2O_4$ nanofibers involve 9.6 at% of Cu ions and 6.8 at% of Zn ions, while $ZnFe_2O_4$ nanofibers contain only O, Fe, and Zn elements. Consequently with increasing zinc content, Cu content is decreases and Zn content is increases gradually.

3.3 Magnetic properties

Tytical hysteresis loops for zinc substituting $CuFe_2O_4$ nanofibers at room temperature are shown in Figure 6. The values of saturation magnetization ($M_s$), coercivity ($H_c$), remanence ($M_r$) and square ratio $M_r/M_s$ are given in Table 1. The magnetic properties of nanofibers vary with changing zinc content. The variation of magnetic properties of $Cu_{1-x}Zn_xFe_2O_4$ nanofibers can be understood in term of cation distribution and exchange interactions between spinel lattices.

The $Cu_{1-x}Zn_xFe_2O_4$ nanofibers with x ≤ 0.6 exhibit ferromagnetic behavior, whereas other nanofibers display paramagnetic character with zero coercivity, zero remanence and non-saturated magnetization. The saturation magnetization initially increases with increasing zinc content to reach a maximum (58.4 emu/g) and then decreases. The increase in saturation magnetization may be attributed to the fact that, small amount of $Zn^{2+}$ ions substituted for $Cu^{2+}$ occupy A sites displacing $Fe^{3+}$ ions from A sites to B sites, which increasing the content of $Fe^{3+}$ ions in B sites. This leads to an increase of magnetic moment in B-site and a decrease of magnetic moment in A-site. So the net magnetization increases, which is consistent with the increase of saturation magnetization. With further increase nonmagnetic $Zn^{2+}$ ions content, an increasing dilution in A sites is present, which results to the collinear ferromagnetic phase breaks down at x = 0.4. For $Cu_{1-x}Zn_xFe_2O_4$ nanofibers (x = 0.6, 0.8, and 1.0), the triangular spin arrangement on B-sites is suitable and this causes a reduction in A-B interaction and an increase of B-B interaction. Therefore, the decrease of saturation magnetization can be explained on the basis of three sublattice Yafet-Kittle model [20].

As shown in Table 1, coercivity ($H_c$) and square ratio ($M_r/M_s$) continuously reduced with increasing $Zn^{2+}$ ions content. These magnetic behaviors of ferrite depend intensely on the spinel structure. For instance, normal spinel ferrite shows an antiferromagnetically ordering, while inverse spinel ferrite shows a ferromagnetic ordering [21]. With increasing $Zn^{2+}$ ions concentration, a transformation from inverse spinel structure of $CuFe_2O_4$ ferrite to normal spinel structure of $ZnFe_2O_4$ ferrite will arises gradually. Consequently, the decrease tendency of $H_c$ and $M_r/M_s$ is directly related to the paramagnetic relaxation effect.

## 4. Conclusion

In conclusion, a series of single-phase $Cu_{1-x}Zn_xFe_2O_4$ (x = 0.0, 0.2, 0.4, 0.6, 0.8 and 1.0) nanofibers were synthesized by electrospinning combined with sol-gel method. The lattice parameter, average grain size, and X-ray density all are found to increase with increasing zinc content. The substitution of zinc ions improves the morphology of Cu-Zn ferrites nanofibers and decreases the diameter of nanofibers. The magnetic properties of $Cu_{1-x}Zn_xFe_2O_4$ nanofibers depend on zinc cotent, which is consistent with the cation distraction and exchange interactions between spinel lattices.

**Acknowledgements**

This paper is supported by National Science Fund of China (11074101), and the Fundamental Research Funds for the Central Universities (860080).

**References**

[1] J. Fang, N. Shama, L.D. Tung, E.Y. Shin, C.J. O'Connor, K.L. Stokes, et al., J. Appl. Phys. 93 (2003) 7483-7485.

[2] Y.C. Wang, J. Ding, J.H. Yin, B.H. Liu, J.B. Yi, S. Yu, J. Appl. Phys. 98 (2005)124306-7.

[3] M.A. Ahmed, N. Okasha, S.I. Ei-Dek, Nanotechnology 19 (2008) 065603-7.

[4] S. Ayyappan, S. Mahadevan, P. Chandramohan, M.P. Srinivasan, J. Philip, B. Raj, J. Phys. Chem. C 114 (2010) 6334-6341.

[5] V. Sepelak, I. Bergmann, A. Feldhoff, P. Heitjans, F. Krumeich, D. Menzel, et al, J. Phys. Chem. C 111 (2005) 5026-5033.

[6] A. Formhals, U.S. Patent No. 1, 975 504 (1934).

[7] D. Li, T Herricks, Y.N. Xia, Appl. Phys. Lett. 83 (2003) 4586-4588.

[8] M. Sangmanee, S. Maensiri, Appl. Phys. A 97 (2009) 167-177.

[9] S. Maensiri, M. Sangmanee, A. Wiengmoon, Nanoscale Res. Lett. 4 (2009) 221-228.

[10] J. Xiang, X.Q. Shen, X.F. Meng, Mater. Chem. Phys. 114 (2009) 362-366.

[11] T. Tsoncheva, E. Manova, N. Velinov, D. Paneva, M. Popova, B. Kunev, et al., Cata. Commun. 12 (2010) 105-109.

[12] S.W. Tao, F. Gao, X.Q. Liu, O.T. Sørensen, Mater. Sci. Eng., B 77 (2000) 172-176.

[13] K. Faungnawakij, Y. Tanaka, N. Shimoda, T. Fukunaga, R. Kikuchi, K. Eguchi, Appl. Catal. B 74 (2007) 144-151.

[14] X. Zuo, A. Yang, C. Vittoria, V.G. Harris, J. Appl. Phys. 99 (2006) 08M909-3.

[15] C.W. Yao, Q.S. Zeng, G.F. Goya, T. Torres, J.F. Liu, J.Z. Jiang, J. Phy. Chem. C 111 (2007) 12274-12278.

[16] X.Q. Shen, J. Xiang, F.Z. Song, M.Q. Liu, Appl. Phys. A 99 (2010) 189-195.

[17] J.H. Nam, Y.H. Joo, J.H. Lee, J.H. Chang, J.H. Cho, M.P. Chun, B.I. Kim, J. Magn. Magn. Mater. 321 (2009) 1389-1392.

[18] J. Xiang, X.Q. Shen, F.Z. Song, M.Q. Liu, Chin. Phys. B 18 (2009) 4960-4965.

[19] M.A. Gabal, Y.M. Al Angari, Mater. Chem. Phys. 118 (2009) 153-160.

[20] M. Ajmal, A. Maqsood, J. Alloy. Comp. 460 (2008) 54-59.

[21] M.H. Khedr, A. A. Farghali, J. Mater. Sci. Technol, 21 (2005) 675-680.

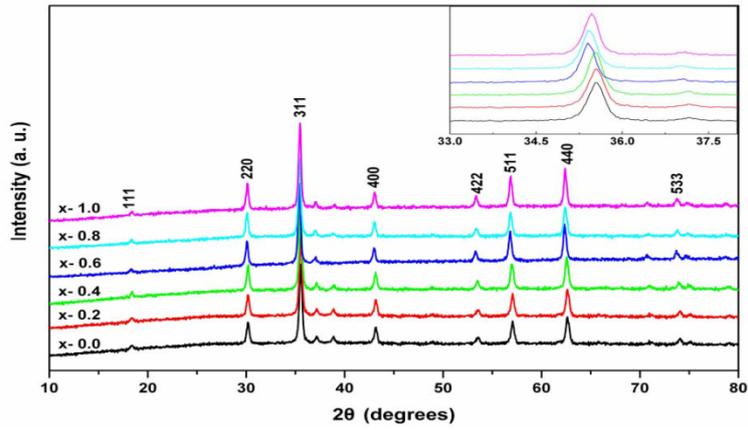

Figure 1. X-ray diffraction patterns of $Cu_{1-x}Zn_xFe_2O_4$ (0.0 ≤ x ≤ 1.0) nanofibers. The inset is XRD patterns from 2θ = 33° to 40°

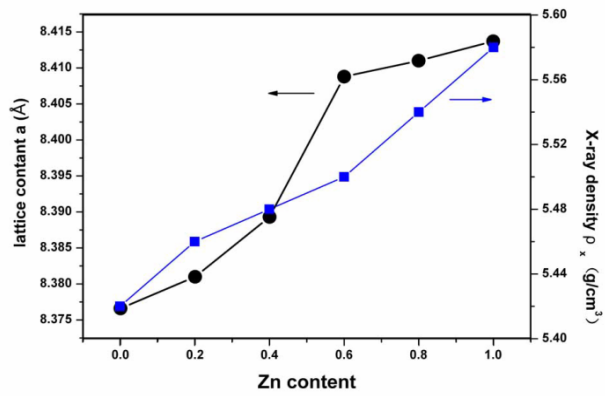

Figure 2. Lattice parameter *a* and X-ray density $\rho_x$ as a function of the zinc content x.

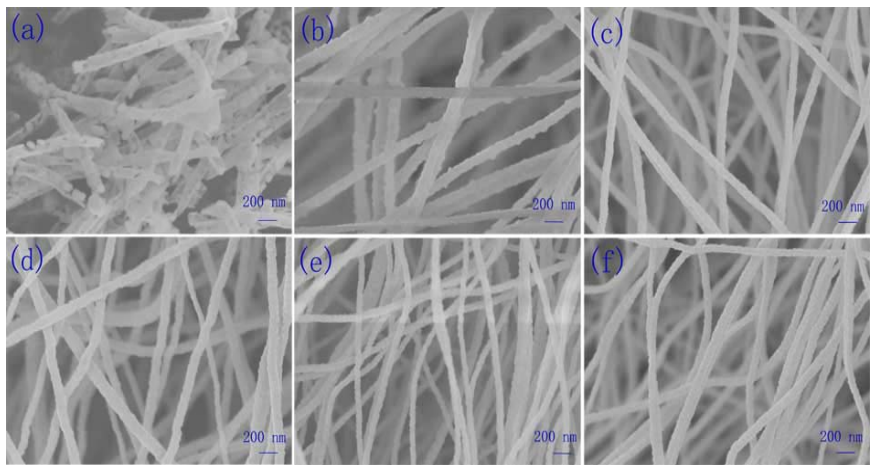

Figure 3. SEM images of $Cu_{1-x}Zn_xFe_2O_4$ nanofibers with different Zn contents (x): (a) 0.0; (b) 0.2; (c) 0.4; (d) 0.6; (e) 0.8 and (f) 1.0.

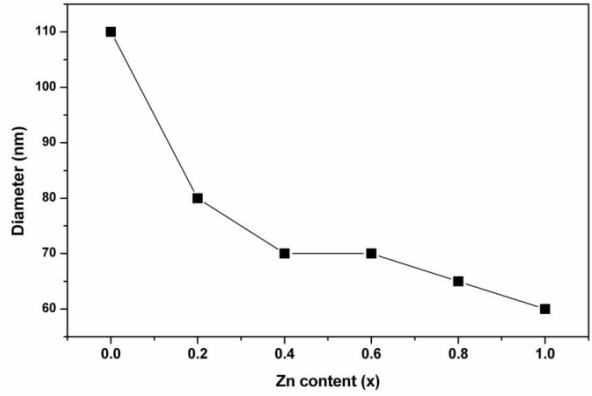

Figure 4. Diameter of $Cu_{1-x}Zn_xFe_2O_4$ (0.0 ≤ x ≤ 1.0) nanofibers depends on the zinc content.

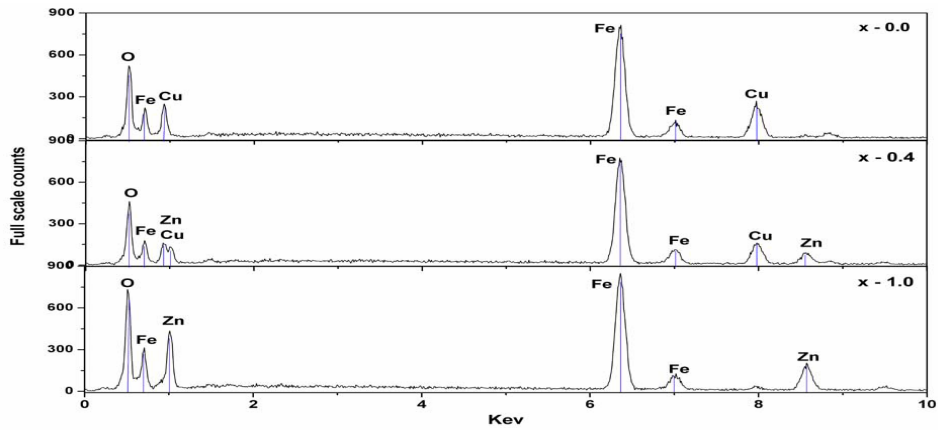

Figure 5. EDS spectra of $Cu_{1-x}Zn_xFe_2O_4$ nanofibers: (a) 0.0; (b) 0.4; and (c) 1.0.

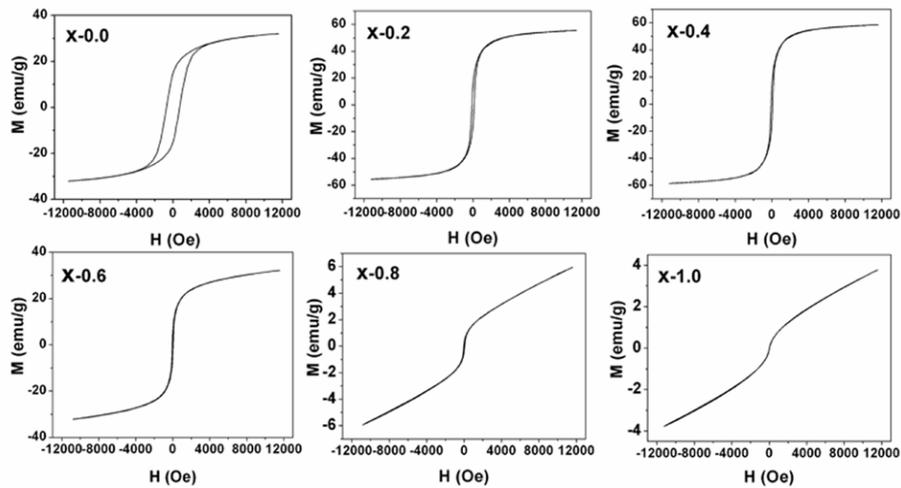

Figure 6. Magnetic hysteresis loops for $Cu_{1-x}Zn_xFe_2O_4$ (0.0 ≤ x ≤ 1.0) nanofibers at room temperature.

Table 1: Table 1. Structural and magnetic parameters for $Cu_{1-x}Zn_xFe_2O_4$ nanofibers system: lattice parameter ($a$), average grain size (D), X-ray density ($\rho_x$), saturation magnetization ($M_s$), coercivity ($H_c$) and square ratio ($M_r/M_s$).

| Sample $Cu_{1-x}Zn_xFe_2O_4$ | Lattice parameter $a$ (Å) | Average grain size D (nm) | X-ray density $\rho_x$ (g/cm$^3$) | $M_s$ (emu/g) | $H_c$ (Oe) | $M_r/M_s$ |
|---|---|---|---|---|---|---|
| x-0.0 | 8.3766 | 26.1 | 5.42 | 31.8 | 723.5 | 0.47 |
| x-0.2 | 8.3810 | 26.9 | 5.46 | 55.3 | 127.8 | 0.28 |
| x-0.4 | 8.3893 | 30.8 | 5.48 | 58.4 | 84.3 | 0.24 |
| x-0.6 | 8.4088 | 31.3 | 5.50 | 32 | 35.2 | 0.13 |
| x-0.8 | 8.4110 | 32.1 | 5.54 | - | - | - |
| x-1.0 | 8.4137 | 30.8 | 5.58 | - | - | - |

Table 2: The elemental analysis results of $Cu_{1-x}Zn_xFe_2O_4$ $(0.0 \leq x \leq 1.0)$ nanofibers by EDS.

| Samples formula $Cu_{1-x}Zn_xFe_2O_4$ | Elemental compositions (at %) | | |
|---|---|---|---|
| | Fe | (1-x) (Cu) | (x)(Zn) |
| $CuFe_2O_4$ | 29.3 | 14.1 | 0.0 |
| $Cu_{0.8}Zn_{0.2}Fe_2O_4$ | 30.0 | 13.2 | 3.8 |
| $Cu_{0.6}Zn_{0.4}Fe_2O_4$ | 29.6 | 9.6 | 6.8 |
| $Cu_{0.4}Zn_{0.6}Fe_2O_4$ | 22.7 | 4.7 | 7.7 |
| $Cu_{0.2}Zn_{0.8}Fe_2O_4$ | 21.7 | 2.8 | 9.1 |
| $ZnFe_2O_4$ | 25.1 | 0.0 | 11.2 |